\documentclass[conference]{IEEEtran}
\IEEEoverridecommandlockouts
\usepackage{booktabs}

\usepackage{cite}
\usepackage{amsmath,amssymb,amsfonts}
\usepackage{algorithmic}
\usepackage{graphicx}
\usepackage{textcomp}
\usepackage{xcolor}
\usepackage{hyperref}
\usepackage{float}
\usepackage{dblfloatfix}
\usepackage{makecell}
\usepackage[normalem]{ulem}
\usepackage{soul}
\usepackage{amsmath}

\usepackage{multirow}
\usepackage{enumitem, array}
\usepackage{caption}
\usepackage{subcaption}
\usepackage[export]{adjustbox}
\usepackage{hyperref}
\usepackage{gensymb}
\usepackage{todonotes}

\graphicspath{{images/}}

\def\BibTeX{{\rm B\kern-.05em{\sc i\kern-.025em b}\kern-.08em
    T\kern-.1667em\lower.7ex\hbox{E}\kern-.125emX}}

\begin{document}

\title{Printing variability of copy detection patterns

\thanks{S. Voloshynovskiy is a corresponding author.}
\thanks{This research was partially funded by the Swiss National Science Foundation SNF No. 200021\_182063.}
}

\author{\IEEEauthorblockN{Roman Chaban, Olga Taran, Joakim Tutt, Yury Belousov, Brian Pulfer, Taras Holotyak and Slava Voloshynovskiy}
\IEEEauthorblockA{Department of Computer Science, University of Geneva, Switzerland \\
\{roman.chaban, olga.taran, joakim.tutt, yury.belousov, brian.pulfer, taras.holotyak, svolos\}@unige.ch}
}
\maketitle

\begin{abstract}

Copy detection pattern (CDP) is a novel solution for products' protection against counterfeiting, which gains its popularity in recent years. CDP attracts the anti-counterfeiting industry due to its numerous benefits in comparison to alternative protection techniques. Besides its attractiveness, there is an essential gap in the fundamental analysis of CDP authentication performance in large-scale industrial applications. It concerns variability of CDP parameters under different production conditions that include a type of printer, substrate, printing resolution, etc. Since digital off-set printing represents great flexibility in terms of product personalized in comparison with traditional off-set printing, it looks very interesting to address the above concerns for digital off-set printers that are used by several companies for the CDP protection of physical objects. In this paper, we thoroughly investigate certain factors impacting CDP. The experimental results obtained during our study reveal some previously unknown results and raise new and even more challenging questions. The results prove that it is a matter of great importance to choose carefully the substrate or printer for CDP production. This paper presents a new dataset produced by two industrial HP Indigo printers. The similarity between printed CDP and the digital templates, from which they have been produced, is chosen as a simple measure in our study. We found several particularities that might be of interest for large-scale industrial applications.

\end{abstract}


\IEEEpeerreviewmaketitle

\section{Introduction} \label{sec:intro}

\begin{figure}
    \centering
    \includegraphics[width=0.4\textwidth]{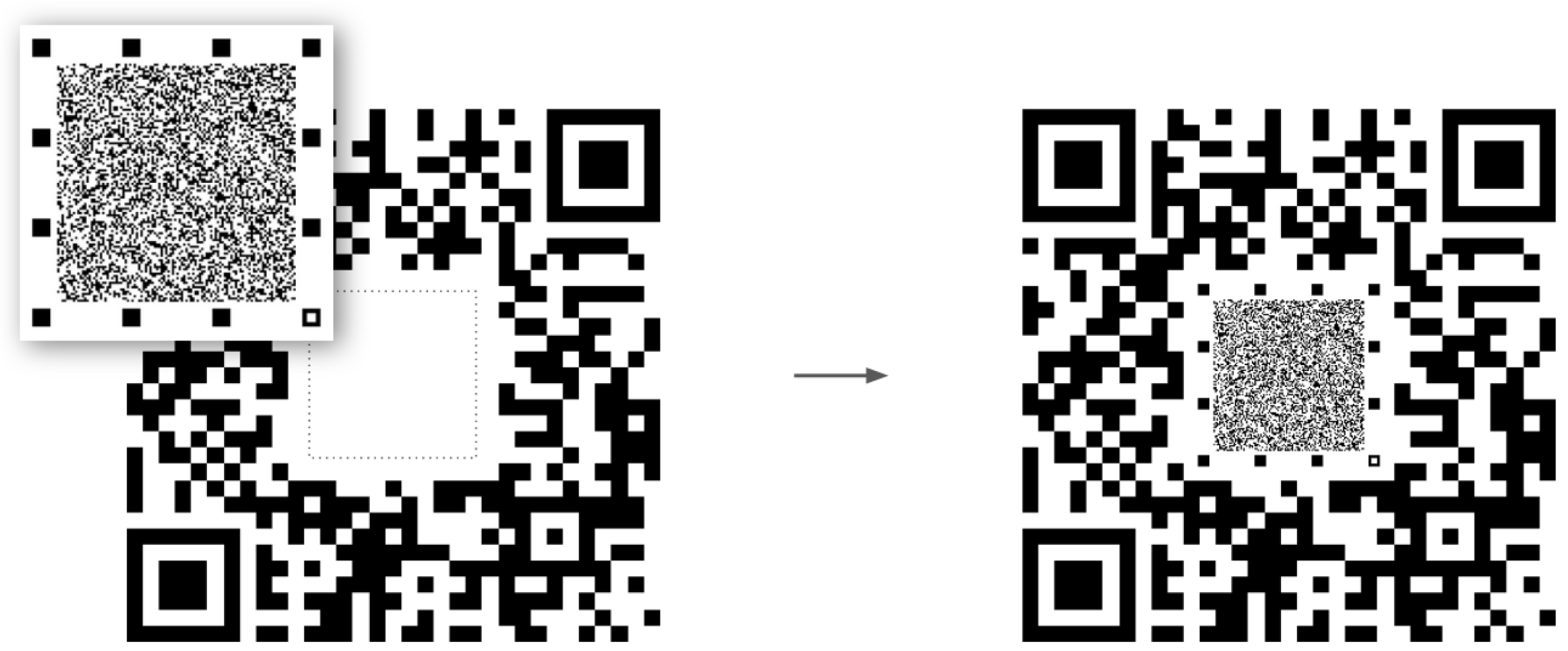}
    \caption{An example of CDP integration into a structure of QR code as suggested by Scantrust \cite{picard2021counterfeit}. CDP is placed inside the QR-code and the copy detection is performed along with the decoding of information stored in the QR-code.}
    \label{fig:cdpqrcode}
    \vspace{-5.5mm}
\end{figure}

In recent years, copy detection patterns (CDP)\cite{picard2004} became an attractive and popular technique for product protection against counterfeiting. CDP are often used for the protection of packaging and security labels. CDP are also used for the protection of pharmaceutical products and vaccines, for example, those against COVID-19 \cite{weforumThisTechnology}. In general, CDP are printed on digital off-set printers but classical off-set and flexo can be used as well. They are easily integrated in the package design as shown in Fig. \ref{fig:cdpqrcode} according to \cite{picard2021counterfeit, weforumThisTechnology }.

Besides the popularity of CDP and its wide usage, there are still some issues that are little studied and raise questions about CDP security in critical applications. This relates to the lack of large-scale public datasets produced on industrial equipment. Moreover, there are numerous researches which aim at challenging CDP security by creating high-quality fakes \cite{9648387} \cite{9648384} \cite{8682967} \cite{10.1145/3335203.3335718}. Such fakes represent a considerable threat to CDP as a protection technology.

In this paper, we aim to fulfilling the gap in the availability of research datasets addressing the variability of CDP and release a new public dataset of CDP printed on two industrial printers HP Indigo 5500 (HPI55) and HP Indigo 7600 (HPI76) under various settings\footnote{The dataset is available \href{https://github.com/sip-group/snf-it-dis/tree/master/datasets/indigo1x1variability}{https://github.com/sip-group/snf-it-dis/tree/master/datasets/indigo1x1variability}.}. The main goal of this study is to investigate the statistical variability of CDP in terms of the deviation of statistics of printed CDP with their digital counterparts in the accordance to the previous researches \cite{picard2008copy}. Each factor of variability is formulated as a separate research question in this work. In summary, we investigated the impact of the following factors:

\begin{itemize}
    \item The difference in CDP statistics produced by two industrial printers HPI55 and HPI76;
    \item The influence of substrate;
    \item The variability in continuous printing and over time;
    \item The effect caused by the acquisition device;
    \item The impact of deviation of printing resolution.
\end{itemize}

Taking into account the mentioned impact factors the main contributions of the paper are twofold:

\begin{itemize}
    \item The new public dataset with synchronized CDP and their digital templates;
    \item The analysis of CDP variability of the above factors.
\end{itemize}

\section{Dataset} \label{sec:data}

In this work, we present a new public CDP dataset specifically created to investigate the CDP variability. The produced dataset of CDP contains digital templates of size 228x228 pixels with 50\% of pixels being white and 50\% black. The digital CDP with the synchronization patterns are allocated on A4 pages. One page contains 12 rows and 12 columns of CDP. The resulting amount of CDP per page is 144. The \textit{basic set} consists of 10 A4 pages with unique CDP codes. This set of 10 A4 pages is then printed with different printing parameters and substrates at different time intervals on two printers. The methodology of our approach for CDP production in details is described in \cite{9648387}.

Taking into account that the methodology of CDP's printing, acquisition, and the following post-processing, i.e., synchronization, grayscale conversion, normalization, etc., is the same as in the previously released Indigo 1x1 base dataset \cite{9648387}, we complement our new dataset by the subset of CDP from the Indigo 1x1 base dataset. To show the importance of the CDP's variability for the authentication, we take into consideration not only the originals but also the ML-based fake CDP from Indigo 1x1 base dataset.

The subsets taken from the Indigo 1x1 base are marked with printing session 0 and contain 720 unique CDP. The goal of this complement is to investigate the printing variability in different time moments while keeping the rest of the parameters fixed.

We investigate the following parameters of variability:

\begin{itemize}
    \item Printers: HPI55 and HPI76.
    \item Printing sessions: 0, 1, 2. Session 0 was printed in May 2021, session 1 in November 2021, and session 2 in February 2022.
    \item Printing substrates: Invercote G (\textit{IG}), Algro Design (\textit{AD}), Atelier (\textit{AT}), Conqueror Stonemarque (\textit{CS}).
    \item Printing resolution: 812.8 and 813 dpi.
    \item Print run: 1 - 10.
    \item Scan run: 1 - 4.
\end{itemize}

\renewcommand{\arraystretch}{1.4}

\begin{table}
    \centering
    \resizebox{.5\textwidth}{!}{%
\begin{tabular}{c|c|c|cccc}
\multirow{2}{*}{\textbf{\begin{tabular}[c]{@{}c@{}}Printing  \\resolution (dpi)\end{tabular}}} &
  \multirow{2}{*}{\textbf{Printer}} &
  \multirow{2}{*}{\textbf{\begin{tabular}[c]{@{}c@{}}Printing  \\session\end{tabular}}} &
  \multicolumn{4}{c}{\textbf{Printing substrate}} \\ \cline{4-7} 
                       &                        &   & \textbf{IG} & \textbf{AD} & \textbf{AT} & \textbf{CS} \\ \hline
\multirow{6}{*}{812.8} & \multirow{3}{*}{HPI55} & 0 & 1 / 1       & -           & -           & -           \\
                       &                        & 1 & 10 / 4      & 1 / 1       & 1 / 1       & 1 / 1       \\
                       &                        & 2 & 1 / 1       & 1 / 1       & 1 / 1       & 1 / 1       \\ \cline{2-7} 
                       & \multirow{3}{*}{HPI76} & 0 & 1 / 1       & -           & -           & -           \\
                       &                        & 1 & 3 / 1       & 1 / 1       & -           & 1 / 1       \\
                       &                        & 2 & 1 / 1       & 1 / 1       & -           & 1 / 1       \\ \hline
\multirow{2}{*}{813}   & HPI55                  & 1 & 3 / 1       & -           & -           & -           \\ \cline{2-7} 
                       & HPI76                  & 1 & 3 / 1       & 1 / 1       & -           & 1 / 1      
\end{tabular}%
}
    \caption{Produced data set: the numbers in cells with the slash indicating "$p/s$" denote $p$ the maximum number of print runs and $s$ the maximum number of scan runs.  }
    \label{tab:setup}
    \vspace{-5mm}
\end{table}

We have chosen two printers to be coherent with our previous studies. We have chosen the mentioned substrates due to their popularity in the production of products' packaging as \textit{IG}, \textit{AD}, and \textit{AT} substrates have roughly the same surface structure, coating, and density. Moreover, we add the substrate \textit{CS} to the analysis as it has a rich texture and complex surface structure.

\begin{figure}[t!]
    \centering
    \includegraphics[width=0.46\textwidth]{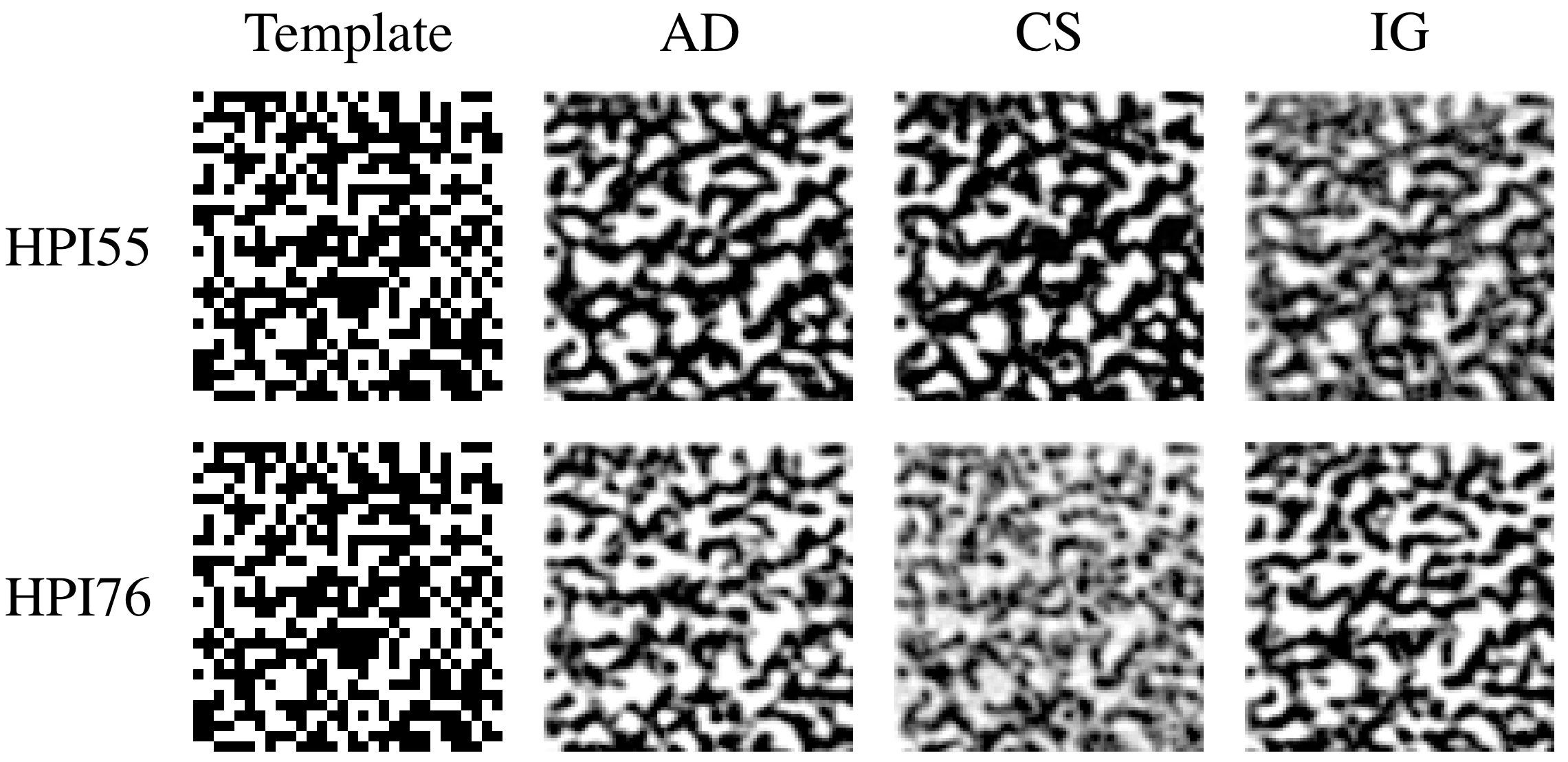}
    \caption{Examples of CDP printed on two different printers and on three different substrates.}
    \label{fig:example1}
    \vspace{-4mm}
\end{figure}

\begin{figure}[t!]
    \centering
    \includegraphics[width=0.46\textwidth]{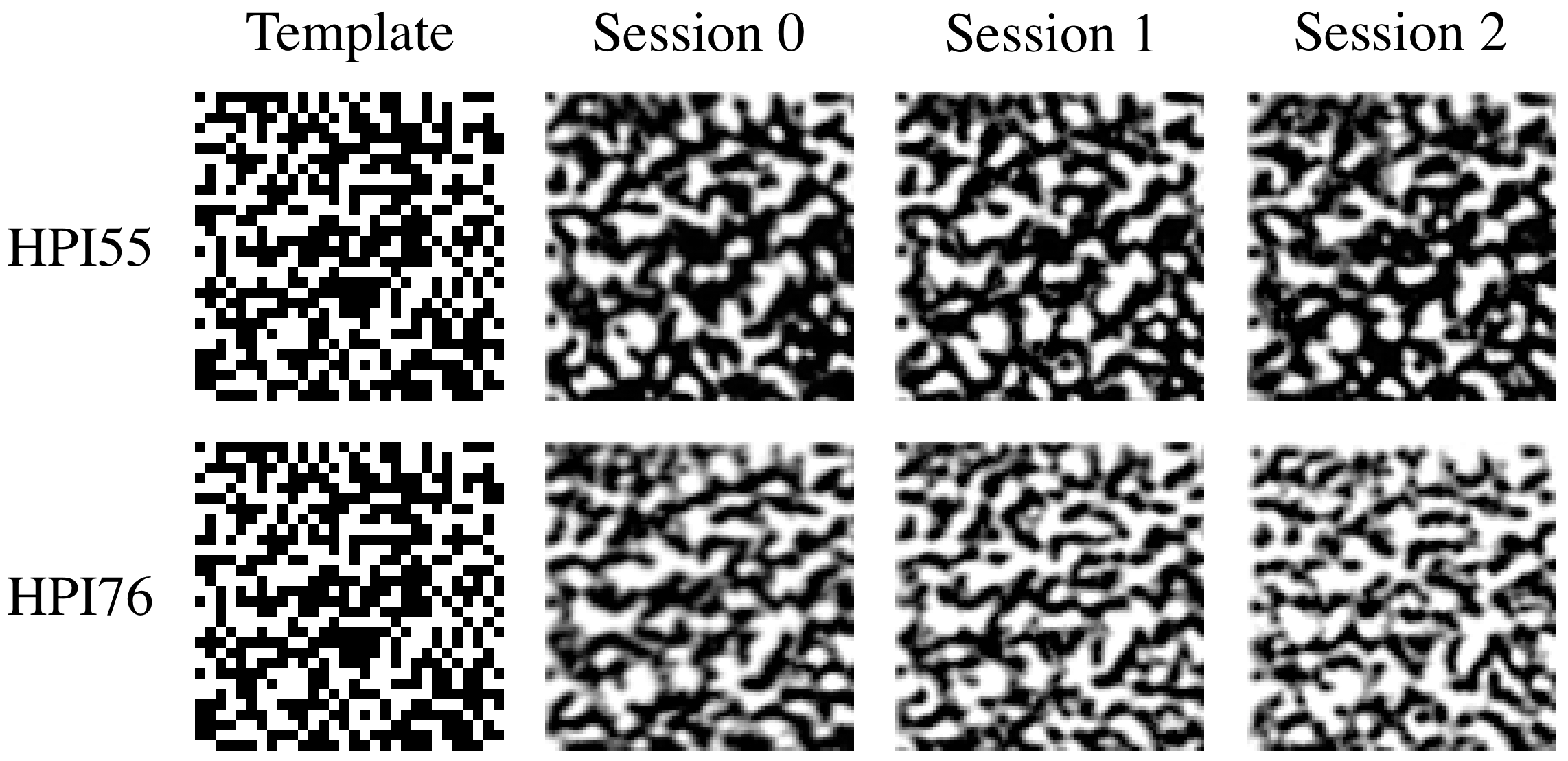}
    \caption{Examples of CDP printed on two different printers on \textit{IG} substrate at three printing sessions.}
    \label{fig:example2}
    \vspace{-4mm}
\end{figure}

\begin{figure}[t!]
    \centering
    \includegraphics[width=0.46\textwidth]{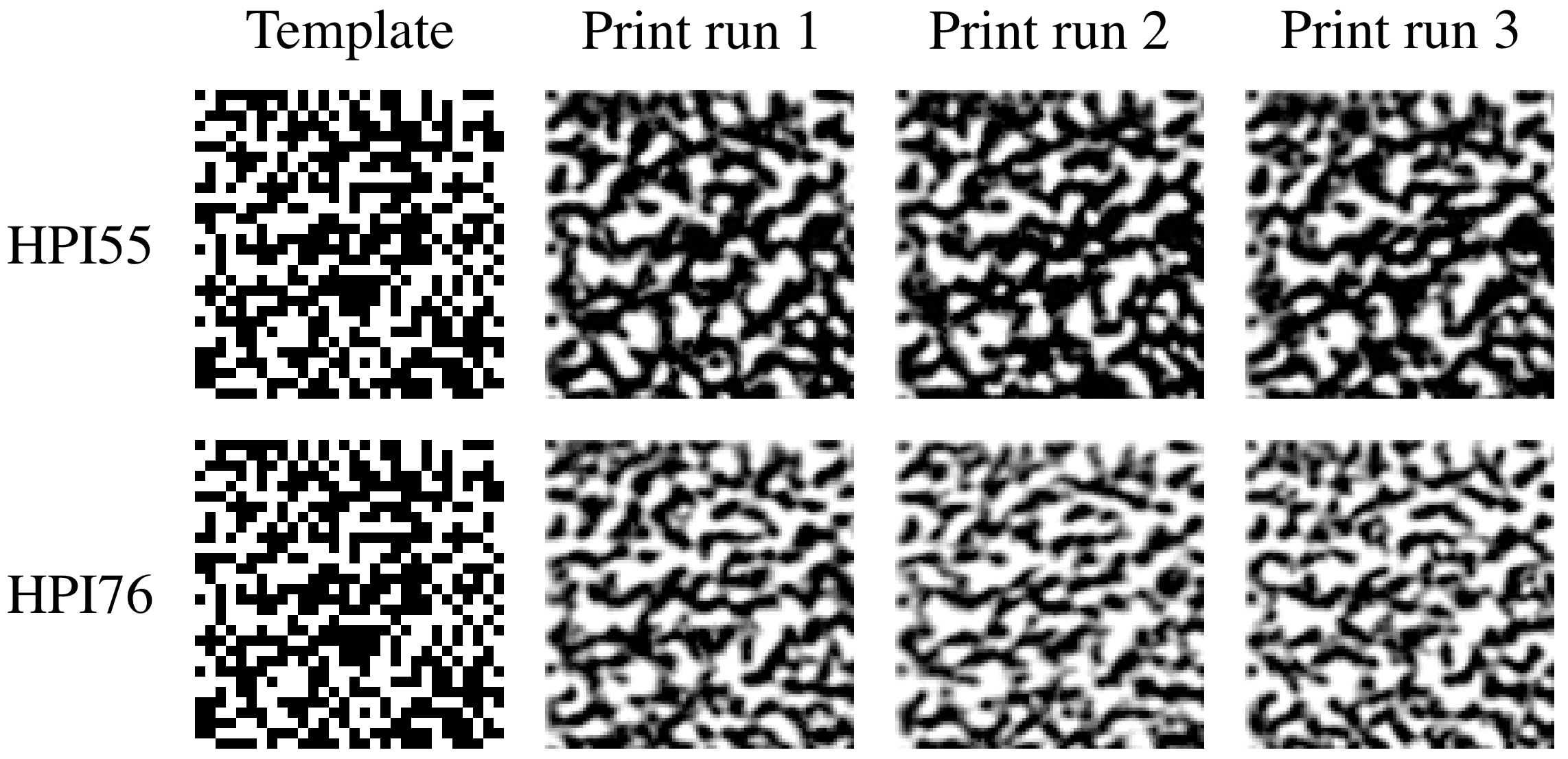}
    \caption{Examples of CDP printed on two different printers on \textit{IG} substrate during the same printing session. Three runs of the same session are shown. One can notice quite essential deviations between three runs.}
    \label{fig:example3}
    \vspace{-4.5mm}
\end{figure}

The overview of the different sets in the produced dataset is presented in Table \ref{tab:setup}. Examples of printed and acquired CDP reflecting the variability due to the impact of the substrate are shown in Fig. \ref{fig:example1}, the impact of time printing sessions for the same substrate in Fig. \ref{fig:example2} and intra-session variability during the continues printing in Fig. \ref{fig:example3}.


\begin{figure*}
    \centering
    \begin{subfigure}{.43\textwidth}
        \centering
        \includegraphics[width=\textwidth]{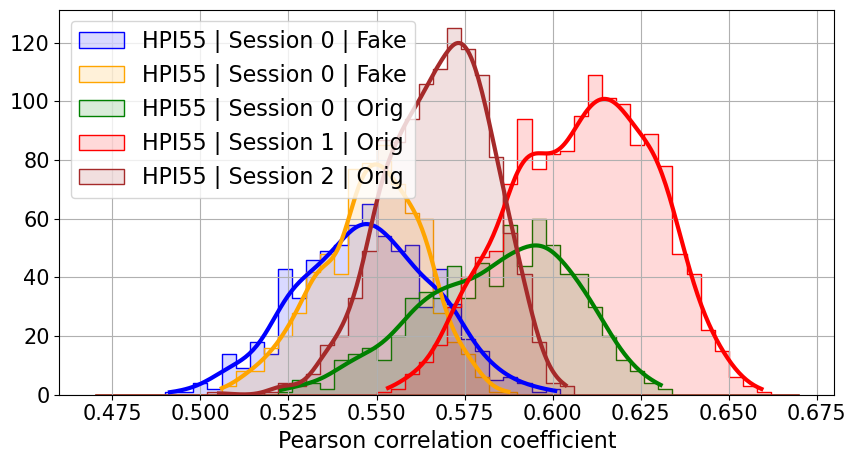}
        \caption{The statistical variability of HPI55.}
        \label{fig:hist_session_printer_55}
    \end{subfigure}
    \begin{subfigure}{.43\textwidth}
        \centering
        \includegraphics[width=\textwidth]{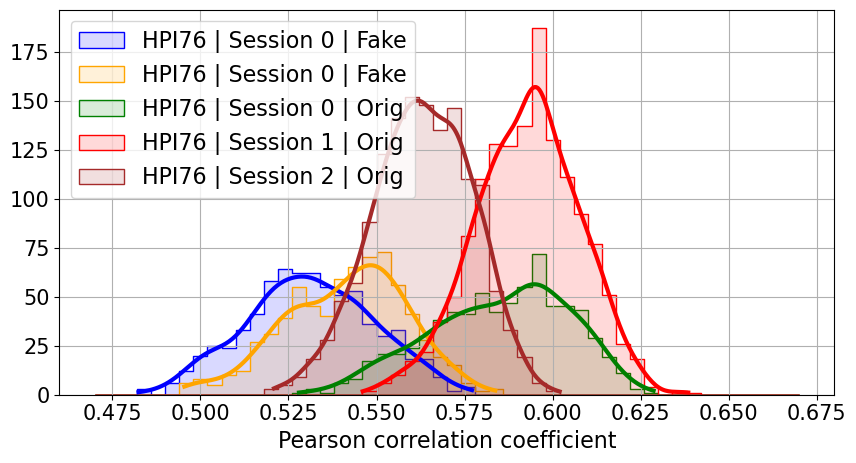}
        \caption{The statistical variability of HPI76.}
        \label{fig:hist_session_printer_76}
    \end{subfigure}
    \caption{The statistical variability over three printing sessions for both printers, including the fakes.}
    \label{fig:hist_session_printer}
    \vspace{-6mm}
\end{figure*}

\section{Investigation of impact factors on the variability of CDP statistics}

Instead of investigating the intra-class variability of CDP statistics, we focus on an authentication setup, where the variability of some similarity score between the printed CDP and its digital reference used for the authentication decision is studied. In this respect, the studied variability is a function of the reference template and a chosen metric. Such an approach has its advantages when directly evaluating the authentication system performance in the space of chosen decision metrics. 

Therefore, after the printing, acquisition, and pre-processing\footnote{The pre-processing includes synchronization of acquired CDP in respect to the digital CDP and min-max normalization.} we calculate a set of similarity metrics for each CDP with respect to the digital template. In our study, we investigate the following similarity metrics: Pearson correlation coefficient (\textit{PCORR}), Hamming distance (\textit{HAMMING}), and structural similarity (\textit{SSIM}) similarly to \cite{1284395}. We provide the results only for \textit{PCORR} and \textit{SSIM} as they are the most discriminative metrics. Additionally, \textit{PCORR} is widely used in CDP-related studies as information-preserving and sufficient statistics under certain assumptions \cite{picard2004} \cite{9648384}.

In our study, we set up a reference set of parameters to be\footnote{Unless it is mentioned otherwise, we use above set of parameters for all figures.}:
\begin{itemize}
    \item Printer: HPI55;
    \item Printing session: 1;
    \item Printing substrate: \textit{IG};
    \item Printing resolution: 812.8 dpi;
    \item Print run: 1;
    \item Scan run: 1.
    \vspace{-0.5mm}
\end{itemize}

\subsection{The variability over three printing sessions}

\begin{figure}
    \includegraphics[width=0.43\textwidth]{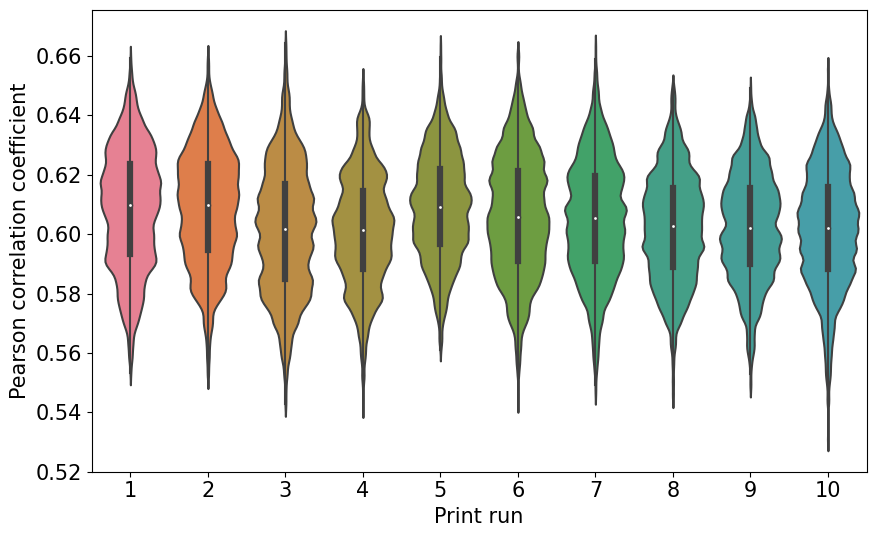}
    \caption{The violin plot of 10 print runs from HPI55 with \textit{PCORR}}
    \label{fig:violin_print_run}
    \vspace{-6mm}
\end{figure}

The printer is a device whose technical conditions and potentially tuning are not stable over time as it requires the replacement of certain parts and inks or toners during the exploitation. Thus, these printer updates can have an impact on the printing quality especially if CDP are produced over a long time interval. Therefore, it is very critical to ensure that the statistics of CDP remain stable over this interval of time. The period can be several months or even a year. To investigate the stability of CDP with respect to the digital templates from which CDP are printed, we considered three printing sessions as described in Section \ref{sec:data} corresponding to May 2021, November 2021, and February 2022. The examples of variabilities are shown in Fig. \ref{fig:example2}. The \textit{PCORR} between the acquired images and their digital counterparts is shown in Fig. \ref{fig:hist_session_printer}. One can observe considerable variability for both printers over three printing sessions. The variability for HPI76 is lower in comparison to HPI55.

The obtained results show that the variability over the time that we investigate expands significantly the distribution of original CDP as shown in Fig. \ref{fig:hist_session_printer} and, hence, increases the probability of false acceptance $P_{fa}$ of fake CDP. At the same time, the distributions of original and fake CDP do not completely overlap. This might help achieve a small enough probability of miss $P_{miss}$ in the considered scenario. However, it is important to investigate the variability of original CDP under different factors that might impact the statistics of printed codes.

It should be pointed out that printers are located in different printing companies and there is a small chance that their maintenance is performed at the same time. Moreover, the micro-climate conditions in the printing companies might impact both printers and substrates. Therefore, it is difficult to establish which factors might impact such considerable variations. In any case, the variability of similarity scores is non-negligible and should be properly taken into account when planning a long-term deployment of CDP-based authentication systems.

\subsection{The variability within one time printing session} \label{sec:printing_variablity}

In this part, we aim at investigating the variability of CDP statistics within one printing session on the extended set. To investigate this variability we printed \textit{basic set} 10 times on HPI55 printer. We summarize the variability per each printing run in Fig. \ref{fig:violin_print_run}. The variabilities over 1 run and overall in 10 runs are relatively small in comparison to the printing in different printing sessions as shown in Fig. \ref{fig:hist_session_printer}.

\begin{figure*}[!t]
    \centering
    \begin{subfigure}{.45\textwidth}
        \centering
        \includegraphics[width=\textwidth]{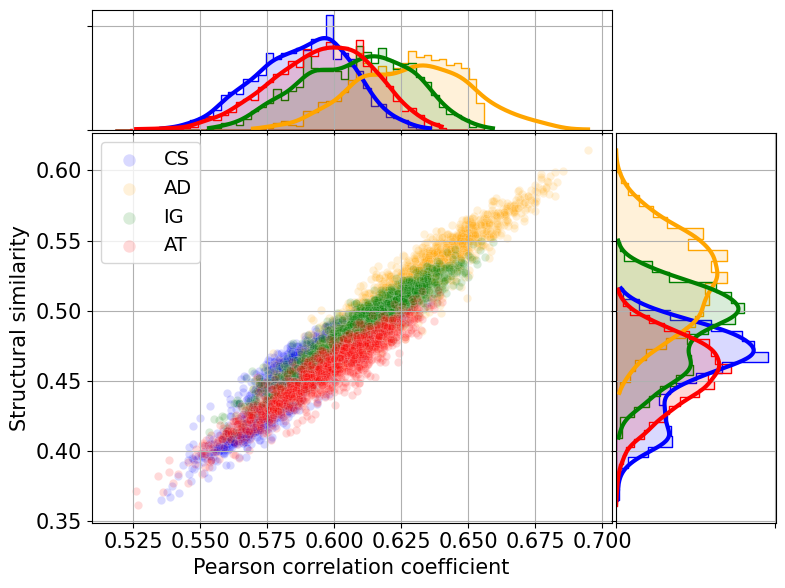}
        \caption{The variability of 4 paper types for HPI55.}
        \label{fig:scatter_paper_55}
    \end{subfigure}
    \begin{subfigure}{.45\textwidth}
        \centering
        \includegraphics[width=\textwidth]{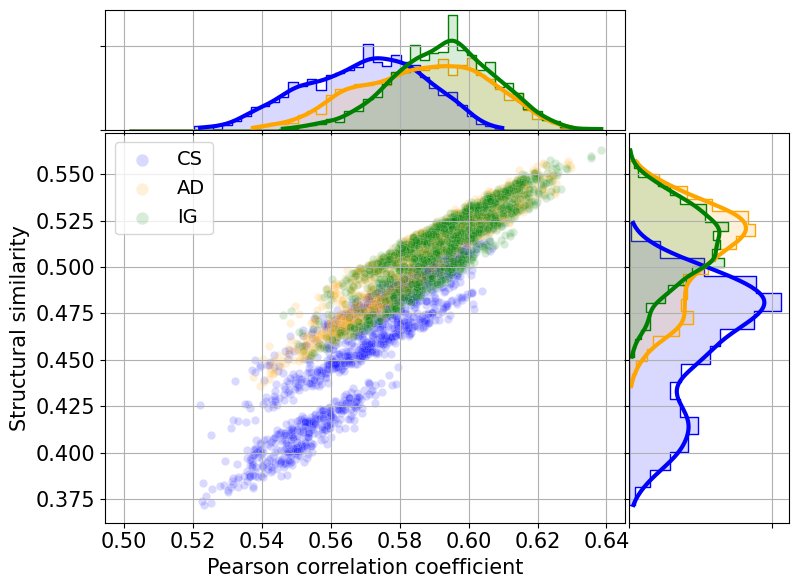}
        \caption{The variability of 3 paper types for HPI76.}
        \label{fig:scatter_paper_76}
    \end{subfigure}
    \caption{The scatter plot with corresponding histograms for \textit{PCORR} (X-axis) and \textit{SSIM} (Y-axis).}
    \label{fig:scatter_paper}
    \vspace{-6mm}
\end{figure*}

It is also important to note that there are some fluctuations between the physical copies printed from the same digital template in the same position on A4 page\footnote{We demonstrate in Section \ref{sec:position} that the position of CDP on an A4 page is also a source of statistical variability due to non-uniform printing in different positions.}. Examples of this variability are shown in Fig. \ref{fig:example3}. The differences between the printed codes are noticeable during visual inspection. However, these fluctuations are relatively small in terms of the chosen similarity measure between CDP and digital templates, i.e., in terms of \textit{PCORR}.

Several factors might explain the variability within one printing session. The first factor is the natural randomness of interaction between the printer toner and substrate leading to the different appearance of CDP printed from the same digital template. Moreover, the quality and uniformity of toner might play an important role. The granularity of toner and the roughness of the substrate might be the factors of impact too. In any case, one can conclude that the printing variability within a single printing session can be neglected for the authentication systems. At the same time, one should not disregard the possibility of further exploring the individual properties of each printed CDP for {\em fingerprinting} purposes \cite{Diephuis:2012:WIFS} that is out of the scope of our present study.

The presented stability of printing in a single printing session can guarantee that printer itself yields "quasi"-deterministic results and we can utilize this for future CDP production.

\begin{figure*}[!t]
    \centering
    \begin{subfigure}{.43\textwidth}
        \includegraphics[width=\textwidth]{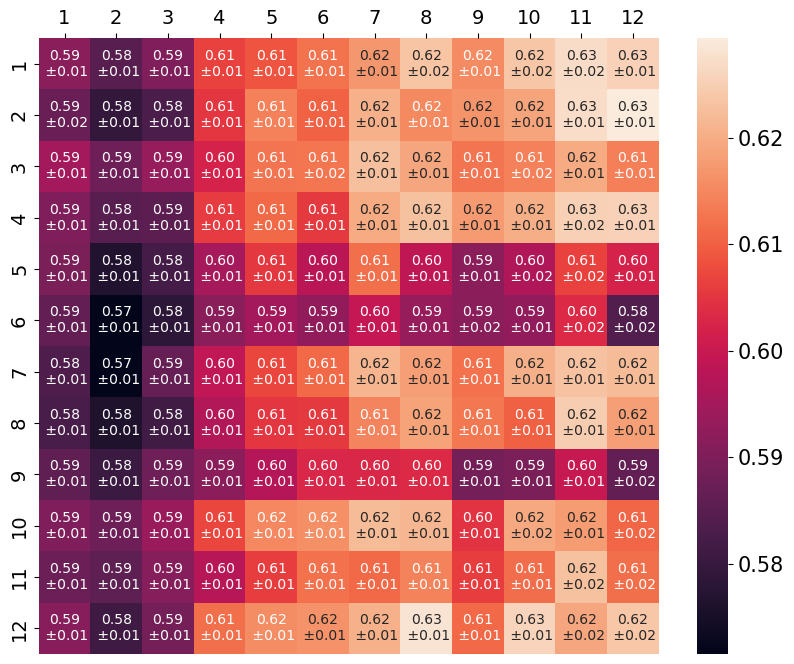}
        \caption{The heat map for HPI55 for 10 print runs.}
        \label{fig:heatmap_print_run_HPI55}
    \end{subfigure}
    \begin{subfigure}{.43\textwidth}
        \includegraphics[width=\textwidth]{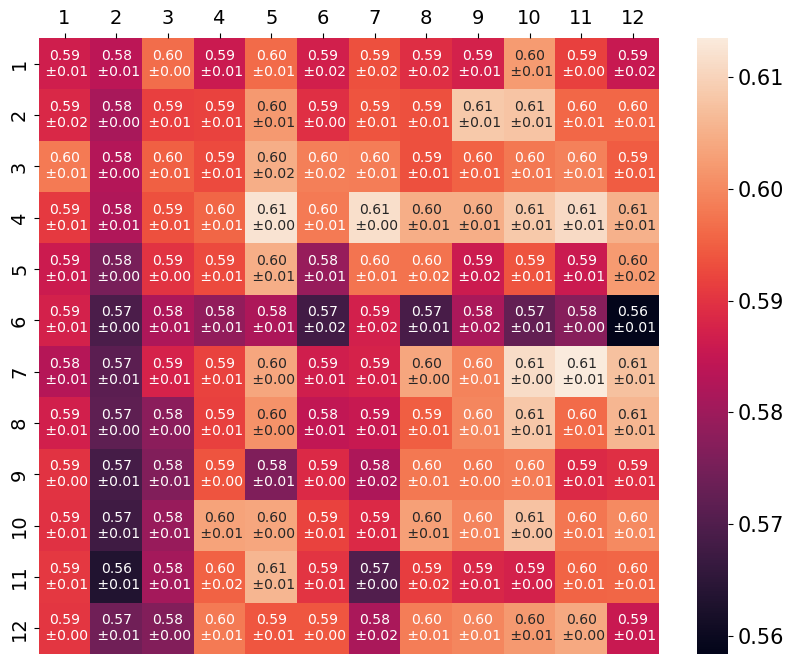}
        \caption{The heat map for HPI76 for 3 print runs.}
        \label{fig:heatmap_print_run_HPI76}
    \end{subfigure}
    \caption{The heat map of CDP similarity score (\textit{PCORR}) per geometrical position on the page averaged over all print runs for the particular printer. Each cell contains mean \textit{PCORR} with its standard deviation.}
    \label{fig:heatmap_print_run}
    \vspace{-6mm}
\end{figure*}

\begin{figure}
    \centering
    \includegraphics[width=0.43\textwidth]{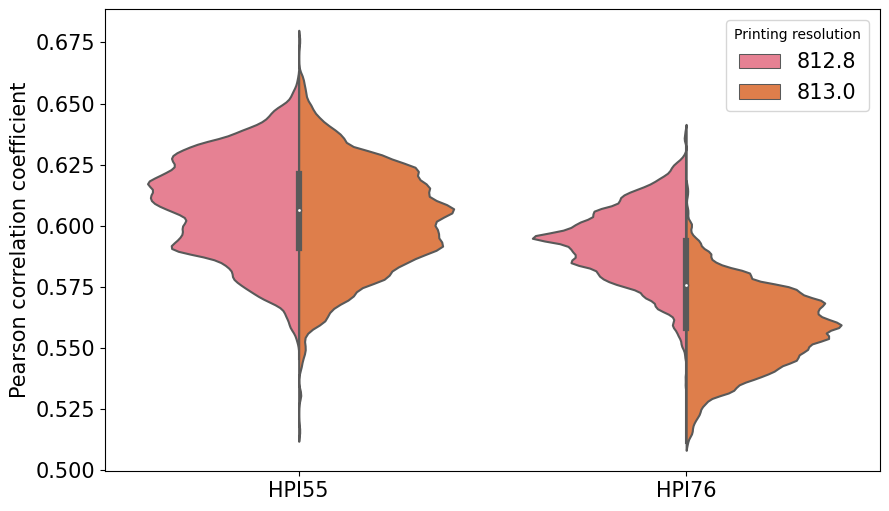}
    \caption{The \textit{PCORR} violin plot for both printers with printing resolution 812.8 and 813 dpi.}
    \label{fig:hist_print_dpi}
    \vspace{-7mm}
\end{figure}

\subsection{The variability due to substrate}

The impact of the substrate on the CDP variability is not well studied and it is often neglected. Sometimes substrates with similar parameters are used interchangeably during the printing session. Thus, our goal is to establish the impact of the substrate on the variability of CDP statistics.

To highlight this impact we use both \textit{PCORR} and \textit{SSIM} to measure the similarity between CDP and digital templates. The obtained results are shown in Fig. \ref{fig:scatter_paper}. First of all, one can observe that the variability of CDP printed on different substrates is approximately the same for HPI55. However, the means of the histograms do not coincide. It is interesting to note that CDP printed on the substrate \textit{AD} provide the largest values of both \textit{PCORR} and \textit{SSIM}, i.e., they are less distorted. The smallest values are obtained for the substrate \textit{CS}. At the same time, we cannot confirm an obvious superiority of \textit{AD} for HPI76. This raises the question of substrate selection for the particular printer. However, we have a stable performance for \textit{IG} for both printers. 
 
It is also interesting to note the statistical behaviors of CDP in a 2D system of coordinates represented by \textit{PCORR} and \textit{SSIM} as shown in Fig.~\ref{fig:scatter_paper}. The substrates have different clustering for the two printers. Considering different clusters for the substrates used for HPI76 printing, one can deploy this property for forensic authentication based on the knowledge of used substrates for the authentic CDP.

The main reason for the obtained results might be explained by the structure and roughness of the paper surface. Due to this, the physical processes vary between various substrate types, because of different toner behaviors on such surfaces. One can conclude that the choice of printing substrate plays quite a critical role in CDP manufacturing.

\subsection{The variability as a function of CDP position} \label{sec:position}

During the experiments, we noticed, that there is some dependency between CDP position on a sheet of A4 paper and its similarity score. We anticipated that some areas of the paper have higher quality CDP reproduction than others. Therefore, our goal is to establish the impact of CDP positioning on the A4 page on the similarity score.

According to Section \ref{sec:data}, we have 12 rows and 12 columns of CDP on the A4 page. For this experiment, we print 100 pages for HPI55 and 30 pages for HPI76. In Fig. \ref{fig:heatmap_print_run_HPI55} we show the similarity scores from all CDP printed over 100 pages in the same position on the A4 page. Thus, each cell contains the average score for 100 CDP located in the same position. The same results for 30 pages of HPI76 are shown in Fig. \ref{fig:heatmap_print_run_HPI76}.

One can observe that the visualized maps possess some regular patterns. For HPI55, columns 1-3 and rows 5, 6, and 9 have a lower \textit{PCORR} than others. On the other hand, HPI76 has another pattern and only column 2 and row 6 stand out among the rest. In terms of numbers, for HPI55 we have the best case which equals 0.63, and the worst 0.57. The difference of 6\% is high enough to be addressed properly in the future.

The observed results can be explained by the same physical properties of the particular printing cylinder and the mechanical feeding elements. The obtained knowledge of the relation between CDP position on the paper might allow us to utilize this information for the filtering of outliers with low values of \textit{PCORR} that is known as a code expurgating technique in information theory \cite{10.5555/1146355}. The codes with the low \textit{PCORR} populate the tail of the distributions and are responsible for the probability of missing. Thus, removing these codes from the analysis might enhance the overall system performance. Alternatively, to minimize the rejection of printed packaging or labels with a low \textit{PCORR} in the discovered positions, one can choose a decision threshold of authentication rule as a function of the code position on the paper. 

\subsection{Impact of printing resolution}

It was considered that chosen printers support different printing resolutions. At the same time, the resolution 812.8 dpi is considered to be a native resolution for both HPI55 and HPI76 as there is no interpolation. However, in some cases, the resolution may be set to a non-native value for some reason. Since setting the printing resolution essentially higher than 812.8 dpi will lead to obvious degradation due to the interpolation, which is also proprietary in the case of HP printers and unknown, we decide to investigate the impact of just a minor deviation from the native resolution. Therefore, we aim at investigating the variability of CDP statistics under the native  812.8 dpi resolution and its proxy 813 dpi resolution.

In Fig. \ref{fig:hist_print_dpi} one can observe histograms for different printing resolutions and printers. For HPI55 there is not any major difference between 812.8 and 813 dpi. The distributions are almost identical and there is no impact on the results. However, for HPI76 we observe a drastic difference: 813 dpi \textit{PCORR} is considerably lower with a mean value lower than 0.05 compared to 812.8 dpi.

\begin{figure*}[!t]
    \centering
    \includegraphics[width=0.95\textwidth]{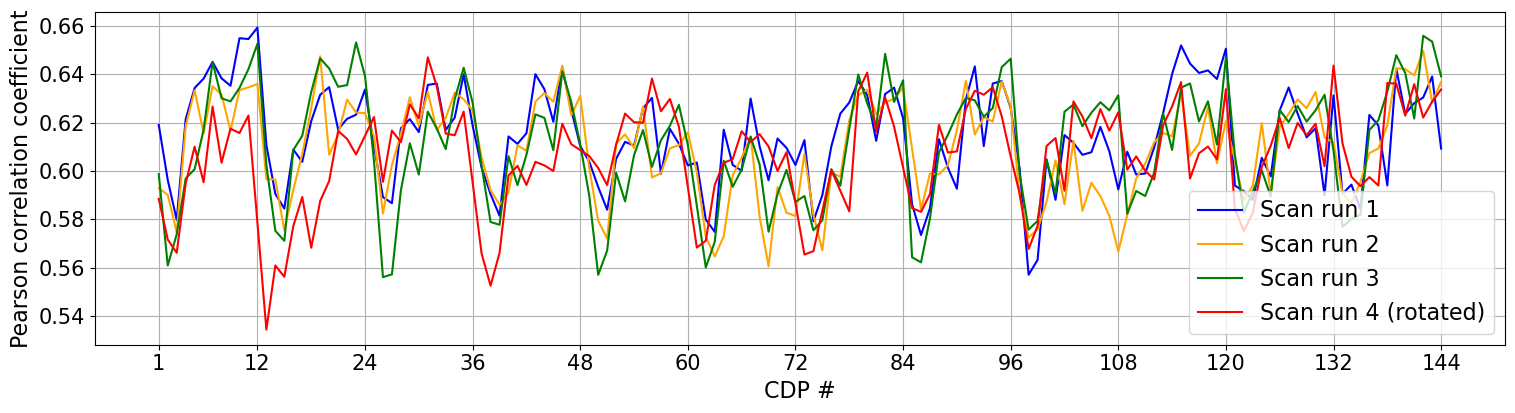}
    \caption{The plot of 4 scan runs for one A4 page. The X-axis is the CDP index, CDP \#12 is located in the 1st row and 12th column, CDP \#47 - is in the 4th row and 11th column.}
    \label{fig:lineplot_scan_run}
    \vspace{-6mm}
\end{figure*}

Such observation indicates significant variability for HPI76 that might be important in practice when the authentication algorithm faces high-quality fakes.

\subsection{Impact of acquisition device}

Along with the analysis of printing variability, it is important to exclude the probability that the acquisition device used in our study is also a source of certain variability. Thus, motivated by the desire to exclude the deviation of acquisition conditions, human factors, and model of acquisition devices, we used the same acquisition device in all our tests. For this purpose we chosen a high-resolution scanner Epson Perfection V850 Pro\footnote{Used software: Epson Scan 2 v6.4.94, resolution = 2400 ppi, unsharp mask: high, brightness = 35, bit depth = 16, color mode = gray. These parameters have been chosen to ensure compatibility with the previous experiments \cite{9648387} where the CDP authentication problem was studied.}. It allows the fast acquisition of a lot of CDP and negates the human factor and lightning conditions, which are present during the acquisition based on a mobile phone. 

The results shown in Section \ref{sec:position} are very crucial since the method of acquisition might be a source of additional variability by itself. Our caveat is that the scanner may be a reason for these results as it is not a perfect acquisition device and has its drawbacks and could be responsible for the correlation between CDP position and its similarity score that should be investigated.

To investigate this problem we perform four scans, three of which were done in the default orientation, meaning that the page was placed on the scanning panel in a particular way and the fourth scan is done with the page rotated by 180\degree. We expected to see that the pattern would change as well. The results are shown in Fig. \ref{fig:lineplot_scan_run} and they prove that our concern is not confirmed. We observe that regardless of the page orientation the acquisition does not impact the resulting similarity score for CDP. The lines on the plot are following the same pattern and present deviations are minor. We see that the red line, for the rotated page, has almost identical behavior.

From the obtained results we can state that scanner does not have an impact on the investigated CDP variability in terms of page placement. However, the impact of other imaging devices such as mobile phones should be additionally investigated.

\section{Conclusions}

In this work, we investigate the variety of factors that influence CDP production and discovered several interesting insights that might have a great impact on understanding the performance of CDP in industrial applications.

We have focused our study on the investigation of CDP variability with respect to their digital templates in a function of printer models, different printing sessions, substrates, which mimic the packaging coating, CDP positioning on the paper, printing resolution, and scanner.

The obtained results show that each of these parameters is worth being taken into consideration in industrial applications. We think that the public availability of the produced datasets of industrially printed CDP under the considered set of parameters will foster a deeper investigation of CDP security to increase the level of trust in this anti-counterfeiting technology for critical applications. 

For future work, we aim at investigating the impact of considered factors for the fake production when the attacker has particular benefits in chosen parameters while the defender does not possess such or vice versa. Moreover, it is also important to investigate the impact of the mobile phone as an imaging device.

\bibliographystyle{IEEEtran}

\bibliography{IEEEexample}

\end{document}